# Parity-time-symmetric closed form optical quadrimer waveguides

Samit Kumar Gupta and Amarendra K. Sarma*
Department of Physics, Indian Institute of Technology Guwahati, Guwahati-781039, Assam, India
*Electronic address: aksarma@iitg.ernet.in

A closed form PT symmetric quadrimer optical waveguide structure, reminiscent of the four-state quantum system found in quantum optics, is studied. The beam dynamics of the structure is studied numerically. The effect of inclusion of nonlinearity and dispersion is also briefly investigated and discussed.

## I. Introduction

Non-Hermitian quantum mechanics based on parity-time (PT) symmetry was proposed by Bender and Boettcher more than a decade ago [1]. Since then, the study of parity-time symmetry have been explored quite extensively in many areas of theoretical physics including quantum field theory [2-6], even though experimental realization of such systems in quantum mechanics is yet to be found. However, recently it has turned out, after the experimental demonstration of PT symmetry in a coupled waveguide system [7-8] and optical mesh lattices [9], that optics may provide enough insight into the mechanism of PT-symmetry along with enormous possible applications [10-15]. In order to appreciate the present work, a brief discussion on the concept of PT symmetry may be useful. Bender and Boettcher have observed in their famous paper [1] that the eigenvalues of a non-Hermitian Hamiltonian (e.g. $\hat{H} = \hat{p}^2/2m + V(\hat{x})$, where m is the mass, V is the complex potential, $\hat{x}$ and $\hat{p}$ are position and momentum operator respectively) may be entirely real if it shares a common set of eigenvectors with the PT-operator, such that $\left[\hat{H}, \hat{P}\hat{T}\right] = 0$ [16]. Here, P is the parity inversion operator, defined as: $\hat{p} \to -\hat{p}, \hat{x} \to -\hat{x}$, and T represents the time reversal operator, defined as: $\hat{p} \to -\hat{p}, \hat{x} \to \hat{x}, i \to -i$. It turns out that the condition for a Hamiltonian to be PT-symmetric is $V(\hat{x}) = V^*(-\hat{x})$ which implies that Re[V(x)]= Re[V(-x)] and Im[V(x)]= -Im[V(-x)]. This concept of PT-symmetry could be extended to optics drawing inspiration from the fact that, the so-called paraxial equation in optics is mathematically isomorphic to that of quantum Schrodinger equation[7]. An optical system is said to exhibit PT-symmetry if the complex refractive-index distribution, $n(x) = n_R(x) + i\, n_I(x)$, of the system satisfies the condition $n_R(x) = n_R(-x)$ and $n_I(x) = -n_I(-x)$. This physically implies that the refractive-index distribution must be an even function of position while the gain/loss profile must be odd. Recently, various linear and nonlinear parity-time symmetric optical structures have been demonstrated theoretically and experimentally. It has been realized that PT symmetry can enable effects, behavior and applications, that would have been impossible in a standard passive optical structure, such as: break-down of the left-right symmetry and power oscillations[17], unidirectional invisibility [10,18], broad-area PT single-mode lasers [13] and coherent perfect absorbers[14,19]. PT-Symmetry has been explored in the context of nonlinear optics also by various research groups. To mention a few: the effect of nonlinearity on beam dynamics in PT-Symmetric potentials [15], stable dark solitons in dual-core waveguides [20], dynamics of a chain of interacting PT-invariant nonlinear dimers [21] and Bragg solitons in nonlinear PT-symmetric periodic potentials [22]. It is interesting to note that even though parity-time symmetry has been investigated in many complex optical structures in various contexts, simple structures like dimer, trimer and quadrimer are still continued to be explored [23-25]. This may be owing to the fact that these PT-symmetric oligomers may act as building blocks for complex PT-symmetric structures or lattices and hence understanding the dynamics of such simple structures may be quite useful for practical applications [26]. In this work, we study a closed form PT-symmetric quadrimer optical waveguide structure. It is appropriate to mention that several authors have studied closed form quadrimer structures in various contexts [27-30].

The beam dynamics of the structure is studied numerically. The effect of inclusion of nonlinearity and dispersion is also briefly investigated and possible practical applications are suggested. The article is structured as follows. In Sec. II the linear PT-symmetric quadrimer structure is presented. In Sec. III the beam dynamics of the structure, with inclusion of nonlinearity and dispersion, is studied using numerical simulations. Finally we draw our conclusions in Sec. IV.

## II. Linear PT-Symmetric Quadrimer

We consider a quadrimer system depicted in Fig. 1.

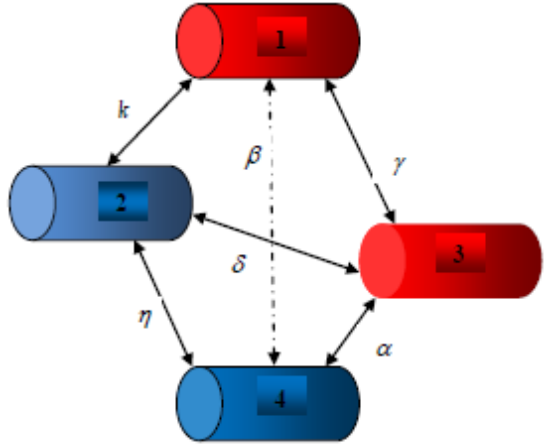

Figure 1. Schematic representation of a quadrimer optical waveguide system. Wave-guides 1 and 3 has gain (+g) while wave guides 2 and 4 has loss (-g). The coupling coefficients between various wave guides are also shown.

The evolution equations describing the optical-field dynamics in the system is given by,

$$\frac{\partial a}{\partial z} = -i H a, \quad (1)$$

where, $a(z) = [a_1(z), a_2(z), a_3(z), a_4(z)]$, with $a_j(z)$ representing the field amplitude in the j-th waveguide and H is the Hamiltonian:

$$H = \begin{bmatrix} ig & -k & -\gamma & -\beta \\ -k & -ig & -\delta & -\eta \\ -\gamma & -\delta & ig & -\alpha \\ -\beta & -\eta & -\alpha & -ig \end{bmatrix} \quad (2)$$

Here, g is the loss/gain coefficient and $[k, \alpha, \eta, \gamma, \delta, \beta]$ are the coupling coefficients between various waveguides as depicted in Fig. 1. The system is PT symmetric if $[H, PT] = 0$, where P is a space-reversal linear operator and T performs element wise complex conjugation [7]. It is straightforward to find that the system is PT-Symmetric under the condition: $k = \alpha$ and $\gamma = \eta$. It is to be noted that, the PT-symmetry of the system is independent of the values of the coupling co-efficient $\delta$ and $\beta$. In rest of the analysis, we ignore the couplings between waveguide 1 and 4 and between 2 and 3, by setting $\delta = \beta = 0$ in Eq. (2). A direct diagonalization of the Hamiltonian gives the following set of four eigenvalues:

$$\lambda = \pm \gamma \pm \sqrt{k^2 - g^2} \tag{3}$$

The eigenvalues are real as long as the gain/loss parameter g is smaller than some critical value, also called the exceptional point[7], $g_{PT} = k$. As the gain/loss parameter g increases above $g_{PT}$, the eigenvalues become complex and the system enters into the so called broken PT-symmetric phase. If the system is kept below the critical point, its four supermodes could be identified as:

$$|1\rangle = \begin{bmatrix} -e^{i\theta} & 1 & e^{i\theta} & -1 \end{bmatrix}^T, \quad |2\rangle = \begin{bmatrix} 1 & e^{i\theta} & -1 & -e^{i\theta} \end{bmatrix}^T$$
$$|3\rangle = \begin{bmatrix} e^{i\theta} & -1 & e^{i\theta} & -1 \end{bmatrix}^T, \quad |4\rangle = \begin{bmatrix} e^{-i\theta} & 1 & e^{-i\theta} & 1 \end{bmatrix}^T \tag{4}$$

where $\theta = \sin^{-1}(g/k)$. None of the supermodes experiences gain/loss and they remain neutral and oscillate during propagation. On the other hand, if the system is kept above the exceptional point, the supermodes are:

$$|1\rangle = \begin{bmatrix} 1 & ie^{-\theta} & -1 & -ie^{-\theta} \end{bmatrix}^T, \quad |2\rangle = \begin{bmatrix} -ie^{-\theta} & 1 & ie^{-\theta} & -1 \end{bmatrix}^T$$
$$|3\rangle = \begin{bmatrix} -ie^{\theta} & 1 & ie^{\theta} & 1 \end{bmatrix}^T, \quad |4\rangle = \begin{bmatrix} 1 & ie^{\theta} & 1 & ie^{\theta} \end{bmatrix}^T \tag{5}$$

where $\cosh(\theta) = g/k$. In this case, the PT-symmetry is spontaneously broken: two of the modes experience amplification while the other two decays exponentially with distance. It is interesting to note that two of the four supermodes above and below the critical point coalesce at the exceptional point, a typical signature of a PT-symmetric system [16].

### III. Beam dynamics: Numerical simulations

In order to understand the beam dynamics of the system, we solve Eq.(1) numerically, subject to the following initial conditions:

$$a_1(z) = 1, a_2(z) = a_3(z) = a_4(z) = 0. \tag{6}$$

The simulation results are depicted in Fig.2. In Fig. 2(a), we depict the power evolutions, in each of the waveguides, for the case of a normal quadrimer system with g=0. On the other hand, Fig. 2(b)-(d) shows the power evolutions of the PT symmetric quadrimer in all the three regimes: below the critical point, at the critical point and above the critical point. Please see the figure caption for the details. One significant effect displayed by PT-symmetric waveguides is the appearance of nonreciprocal wave propagation [8]. In fact, the structure considered in this work also exhibits nonreciprocity, even in the presence of nonlinearity and dispersion. Our numerical investigation shows that the beam propagation is quite sensitive to the initial condition.

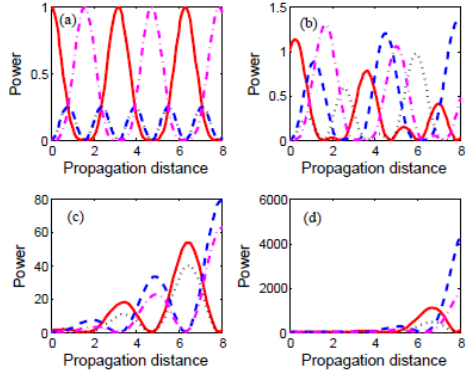

Figure 2. Spatial evolution of power in each site of the quadrimer with $k=1$ and $\gamma=1$. The solid red curve refers to site 1, dotted black refers to site 2 and dashed blue refers to site 3 while dashdot magenta corresponds to site 4. (a) normal quadrimer with $g=0$ (b) $g=0.5$, below the critical point (c) $g=1$, at the critical point and (d) $g=1.1$, above the critical point.

### III a. Role of nonlinearity

The role of nonlinearity on the propagation dynamics may be investigated very quickly through direct numerical simulations of the following coupled equations:

$$\frac{\partial a_1}{\partial z} = i k a_2 + i \gamma a_3 + g a_1 + i |a_1|^2 a_1$$

$$\frac{\partial a_2}{\partial z} = i k a_1 + i \gamma a_4 - g a_2 + i |a_2|^2 a_2$$

$$\frac{\partial a_3}{\partial z} = i k a_4 + i \gamma a_1 + g a_3 + i |a_3|^2 a_3$$

$$\frac{\partial a_4}{\partial z} = i k a_3 + i \gamma a_2 - g a_4 + i |a_4|^2 a_4$$

(7)

Results of direct numerical integrations, subject to the initial conditions (6), is shown in Fig. 3.

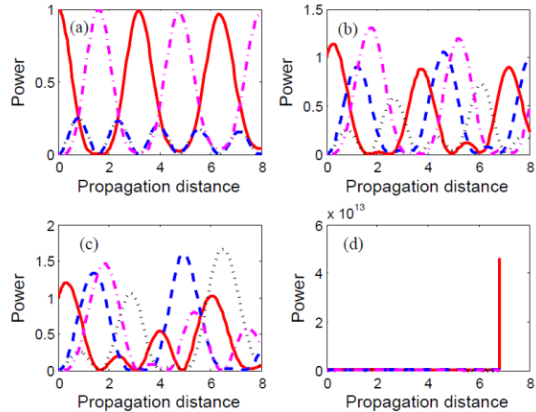

Figure 3. Spatial evolution of power in each site of the quadrimer with $k=1$ and $\gamma=1$ in the presence of nonlinearity. The solid red curve refers to site 1, dotted black refers to site 2 and dashed blue refers to site 3 while dashdot magenta corresponds to site 4. (a) normal quadrimer with $g=0$ (b) (c) $g=0.6$ and (d) $g=0.7$.

The presence of nonlinearity shifts the critical point of the system. As could be seen from Fig. 3(d), as the gain/loss parameter 'g' increases, at a certain value of the parameter there is an abrupt

PT phase transition and power keeps on building exponentially in one of the waveguides. (For the parameters chosen for the current simulation, this value is nearly 0.7, i.e. $g_{cr} \approx 0.7$) This could be understood easily using the source-and-sink model of Ref.[31]. If the coupling of the waveguides is sufficiently strong, then the system is in equilibrium and one can could observe Rabi-like oscillation. On the other hand, when the coupling becomes too weak, the Rabi oscillations ceases and the system is no longer able to maintain equilibrium resulting in exponential increase in power in one of the waveguides and exponential decrease in others.

### III b. Role of dispersion

It may be useful to study the effect of dispersion on the beam dynamics of the quadrimer structure. To do, we consider Eq. (7) without the nonlinear terms and including the so called group velocity dispersion term.

$$\frac{\partial a_1}{\partial z} = i k\, a_2 + i \gamma a_3 + g\, a_1 + \frac{i}{2}\sigma_1 \frac{\partial^2 a_1}{\partial \tau^2}$$

$$\frac{\partial a_2}{\partial z} = i k\, a_1 + i \gamma a_4 - g\, a_2 + \frac{i}{2}\sigma_2 \frac{\partial^2 a_2}{\partial \tau^2}$$

$$\frac{\partial a_3}{\partial z} = i k\, a_4 + i \gamma a_1 + g\, a_3 + \frac{i}{2}\sigma_3 \frac{\partial^2 a_3}{\partial \tau^2} \qquad (8)$$

$$\frac{\partial a_4}{\partial z} = i k\, a_3 + i \gamma a_2 - g\, a_4 + \frac{i}{2}\sigma_4 \frac{\partial^2 a_4}{\partial \tau^2}$$

Here $\sigma_i$ is the so called group velocity dispersion (GVD) parameter in the $i$-th waveguide. Eq. (8) could also be written in the frequency domain, by defining the Fourier transform, $\tilde{a}_i(z,\omega) = \int_{-\infty}^{\infty} d\tau \exp(i\omega\tau) a_i(z,\tau)$. Now, we would like to consider the following three cases which we find particularly interesting:

**Case 1:** Sites have equal dispersion

Let the magnitude of the GVD parameter to be the same in all the waveguides. We can easily find the set of four eigenvalues of the corresponding Hamiltonian to be: $\tilde{\lambda} = \alpha \pm \gamma \mp \sqrt{k^2 - g^2}$, where $\alpha = (|\sigma|\omega^2 / 2)$. It is worthwhile to note that if dispersion is taken into account, the PT symmetry threshold of the system remains unaffected, irrespective of the sign of the parameter $\sigma$. A pulse launched at a site, as expected, gets broadened during its evolution. This is illustrated in Fig. 4, where a Gaussian pulse is launched at site 1 of the quadrimer.

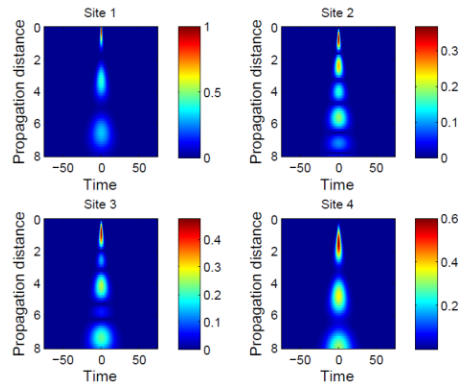

Figure 4. Contour plots for evolution of intensity in each site of the dispersive quadrimer

**Case 2:** Sites have alternate dispersion

If the waveguides are chosen to have alternate dispersion, say site 1 has normal dispersion ($+\sigma$) while site 2 has anomalous dispersion ($-\sigma$) and so on, then system has the following set of eigenvalues:

$$\tilde{\lambda} = \pm\gamma \pm \sqrt{k^2 + (\alpha + ig)^2} \qquad (9)$$

Clearly, the eigenvalues are purely complex and the quadrimer structure do not exhibit PT symmetry at all.

**Case 3:** Site 1 and 2 has normal dispersion while site 3 and 4 has anomalous dispersion or vice-versa.

In this case the system has the following set of eigenvalues:

$$\tilde{\lambda} = \pm\sqrt{\alpha^2 - g^2 + k^2 + \gamma^2 \mp 2\sqrt{(k^2 - g^2)(\alpha^2 + \gamma^2)}} \qquad (10)$$

It is easy to see that, here, eigenvalues become purely real for $k = g$, on the other hand if $k < g$ the eigenvalues become complex. Hence we may conclude that the dispersive quadrimer of case 2 is PT symmetric for $g < g_{PT} = k$. Moreover, it seems dispersion may play a fundamental role in the beam dynamics of the structure.

It may be interesting to investigate the combined effect of both nonlinearity and dispersion on the evolution of an optical pulse. To do so, we add the so-called anomalous group velocity dispersion term, in normalized units, to Eq. (7), thereby making it a system of coupled nonlinear Schrodinger equation (CNLSE) [32]. Then, the set of equations are solved numerically by launching a fundamental soliton in site 1, keeping rest of the sites empty. Fig. 5 depicts the simulation results. At the outset we simulate the CNLSE with $g = 0.7$. Fig. 5(a) plots the output power of the fields. It could be seen that the PT-phase transition is avoided with the introduction of dispersion into the system. Fig. 5(b) depicts the output profiles of the fields with g=0.3. The corresponding contour plots for spatio-temporal evolution is shown in Fig. 6. It can been that the soliton launched initially in site 1 is getting switched to site 4 at the end of the given propagation distance. In fact with judicious choice of the propagation length of the quadrimer, one can obtain the soliton at the desired site.

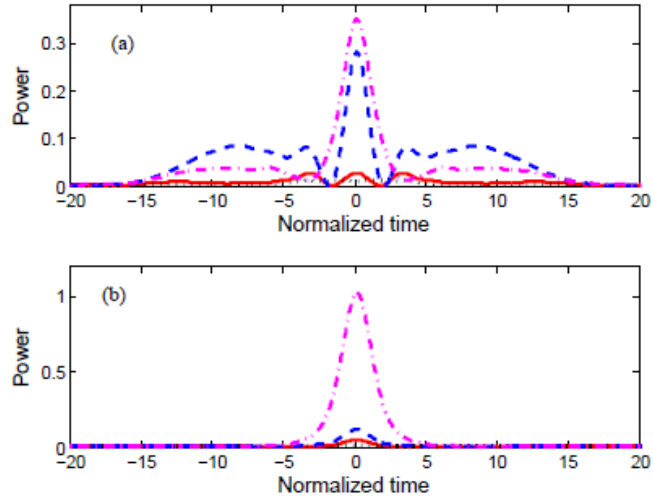

Figure 5. Output power in each site of the quadrimer with $k = 1$ and $\gamma = 1$ in the presence of nonlinearity and dispersion. The solid red curve refers to site 1, dotted black refers to site 2 and dashed blue refers to site 3 while dashdot magenta corresponds to site 4. (a) $g = 0.7$ and (b) $g = 0.3$.

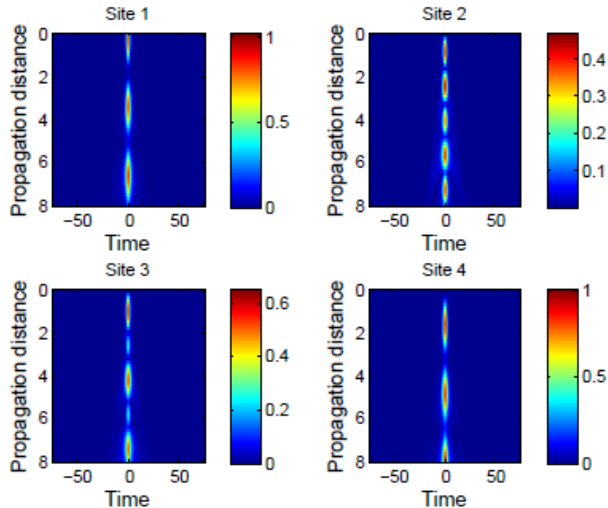

Figure 6. Contour plots for evolution of intensity in each site of the quadrimer corresponding to the case in Fig. 5(b).

The PT-Symmetric quadrimer considered in this work may be exploited for various applications such as tunable $4\times 4$ spatial optical switch [33] and soliton switching [34]. Recently the stability of solitons in a dimer have been investigated [35] and similar analysis may be carried out for the case of a quadrimer also. However, we have checked the stability of solitons in the quadrimer system numerically and find that the soliton is stable as long as the order of the soliton, N, lies in the range: $0.85 < N < 1.2$. It may be worthwhile to mention that our numerical simulation (not shown here) also confirms the nonreciprocal soliton evolution in the PT- symmetric nonlinear quadrimer.

## IV. Conclusions

In the present work we have studied a closed form PT-symmetric quadrimer optical waveguide structure. The beam dynamics of the structure is studied numerically. The effect of inclusion of nonlinearity and dispersion is also briefly investigated and discussed. The structure considered is reminiscent to the four-state quantum system found in quantum optics [36]. It is anticipated that, considering the recent interest in quantum-optical analogies [37] and PT-symmetry in atomic medium [38], careful study of such structures might provide lots of insight into the dynamics of more complex configurations or structures.